\documentclass[%
aip,
 sd,%
 amsmath,amssymb,
 reprint,%
]{revtex4-1}

\usepackage{graphicx}
\usepackage{dcolumn}
\usepackage{bm}
\usepackage{CJK}
\draft 
\usepackage{graphicx}
\usepackage{epsfig}
\usepackage{epstopdf}

\begin{document}


\title{Quantum Bit Commitment Combining with BB84 Protocol} 



\author{Linxi Zhang}
\email{zhanglinxi@xidian.edu.cn}
\affiliation{State Key Laboratory of Integrated Services Networks, Xidian University, Xi¡¯an 710071, China}

\author{Changhua Zhu}
\affiliation{State Key Laboratory of Integrated Services Networks, Xidian University, Xi¡¯an 710071, China}
\author{Nan Zhao}
\affiliation{State Key Laboratory of Integrated Services Networks, Xidian University, Xi¡¯an 710071, China}
\author{Changxing Pei}
\affiliation{State Key Laboratory of Integrated Services Networks, Xidian University, Xi¡¯an 710071, China}


\date{\today}

\begin{abstract}
We proposed a new quantum bit commitment scheme in which secret key need not to be provided by other quantum key distribution system. We can get the bit commitment with probability p by adding a waiting time in a frame during operating the BB84 protocol. Then the measurement outcomes can be encrypted by one-time pad with the key generated by BB84 protocol. We can also obtain the redundant secret keys to encrypt other information in this quantum key distribution system. The new scheme can be used to perform routing operation by which the source can be made sure that the virtual circuits it chosen doesn¡¯t influence on other relay routes in a large-scale trusted relay-based quantum network.
\end{abstract}


\maketitle 


\section{Introduction}
Bit commitment is one of the fundamental cryptographic protocols which was first proposed by Blum \cite{Blum81}. It is a cryptographic protocol between two mistrustful parties conventionally called Alice and Bob. Bit commitment usually consists of two phases. In commitment phase, Alice sends a bit b which she wants to commit to the Bob. It is concealing if Bob cannot know b before Alice unveils it. In the unveiling phase, Alice unveils the bit b to Bob. It is binding if Alice cannot change b once she has committed to it. Bit commitment can be used as a block of a wide range of other cryptographic protocols, such as coin tossing \cite{Blum82}, zero-knowledge proofs \cite{Brassard88}, oblivious transfer and secure two-party computation \cite{kilian91}.

In classical bit commitment protocol, it is accomplished by using computational complexity assumptions to ensure the security. When it comes to the quantum computer, this security does no longer exist. In 1993, a protocol was proposed to realize bit commitment in the framework of quantum mechanics. In 1997, Dominic Mayers have proved that the unconditionally secure of the quantum bit commitment is impossible \cite{Mayers97}. And at the same year, Hoi-Kwong Lo and H. F. Chau have also shown that non-relativistic quantum bit commitment is insecure because of the Alice¡¯s cheating by using an Einstein-Podolsky-Rosen type of attack \cite{Kwong97}$^,$\cite{Kwong98}.

As the increasing consideration of the relativistic quantum information, Adrian Kent proposed a new unconditionally secure quantum bit commitment scheme by transmitting measurement outcomes \cite{Kent99}$^,$\cite{Croke2012}, based on Minkowski causality \cite{Kent2012} and the properties of quantum information \cite{Kent2013}$^,$\cite{Palmer2013}. Then more and more experiments were proposed to prove the feasibility of this method \cite{Liu2014}$^,$\cite{Lunghi}.
\section{P-bit commitment scheme}
We find that Adrian Kent¡¯s quantum bit commitment scheme needs not only to measure the signals received, but also to use other quantum key distribution systems to transmit some of its outcomes. According to the quantum bit commitment protocol and the BB84 protocol, we both need signal photon detectors which is huge waste of resources. Moreover, we need such a lot of secret keys generated by the BB84 protocol to encrypt the measurement outcomes with one time pad that we cannot use a same synchronous clock in both quantum bit commitment protocol and BB84 protocol. We assume that Alice has 4N measurement outcomes in a frame. The generation key rate r,the channel error tolerance, $Q_{tol}$, the required correctness, $\epsilon_{cor}$, the error correction leakage, $leak_{EC}$, the protocol  is $\epsilon_{sec}$ secret if the generation key rate r satisfies\cite{Scarani2008}$^,$\cite{Tomamichel}:
\begin{equation}
r\leq1-h(Q_{tol})-\frac{leak_{EC}}{4N}-\frac{1}{4N}\log\frac{2}{\epsilon_{sec}^2\epsilon_{cor}}
\end{equation}
$h(\cdot)$ denotes the binary Shannon entropy, $h(x)=-x\log x-(1-x)\log (1-x)$.

We can simplify the inequality above. First, we assume that we have perfectly state preparation. So we can get the length of the raw key is $R=4N*(1-h(Q_{tol} ))$. In error correction, for the purpose of the optimization, we assume an error correction leakage of $leak_{EC}=4N*f(Q_{tol} )*h(Q_{tol} )$ with $f(Q_{tol} )\approx1.2$. In privacy amplification, we discard the length of the key is:
\begin{equation}
R*h(\frac{Q_{tol}}{1-h(Q_{tol})})\approx4N*h(Q_{tol})*(1-h(Q_{tol}))
\end{equation}

To sum up, we can get the final key rate r satisfies:
\begin{gather}
r=1-h(Q_{tol})-1.2(Q_{tol})-h(Q_{tol})*(1-h(Q_{tol})) \notag \\
=(1-h(Q_{tol}))^2-1.2h(Q_{tol})
\end{gather}

Alice chooses random basis to measure the incoming signals. So when $N\rightarrow\infty$, there are nearly 2N measurement outcomes measured in the same basis from 4N incoming signals which has been measured in a frame. These nearly 2N measurement outcomes need to be encrypted by quantum key distribution with one time pad. So if we do not depend on the other quantum key distribution systems, the relationship between the generation rate and the measurement outcomes satisfies:
\begin{equation}
\frac{1}{2}*4N*r\geq2N
\end{equation}

We cannot reach this generation rate $r\geq1$ if we use the same synchronous clock. Otherwise, we need to be built two independent systems that costs more waste of resources.

 So we propose a new quantum bit commitment scheme combining with BB84 protocol and it can be utilized in the large-scale relay network routing method. Alice commits a bit b by choosing form 4N measurement outcomes transmitted from Bob in the same basis in a frame. In our scheme, we only pay attention on the basis which Alice chooses split in 2N rectilinear basis and 2N diagonal basis and abort other conditions. In 2N measurement outcomes she chooses, we pick the number of x sequences to be the commitment sequences and encrypt these sequences in another way $(x\leq C_{2N}^N )$. From this way, we can get the probability $p=\frac{xC_{4N}^{2N}}{2^{4N+2N}}$  to commit the bit b. Then Alice transmits the sequences to the other two trustful relays nearby encrypted with one time pad by using the secret these new scheme generated. As the other two relays both receiving the last outcomes from Alice, they will wait for a moment, and then to verify whether the outcomes are consistent.

\begin{figure}
 \centering
 \includegraphics [width=0.4\textwidth,height=0.4\textheight]{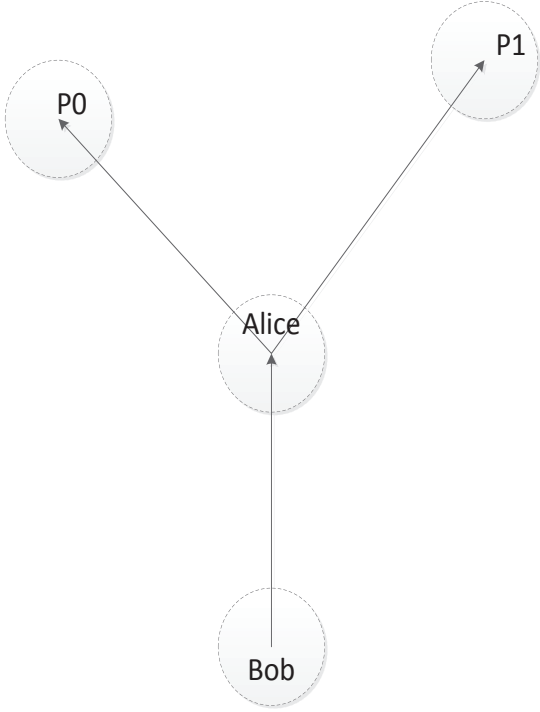}
 \caption{ It is a part of a large-scale quantum relay network. Different from A. kent¡¯s model, we denote that $P_0$ and $P_1$ are both Alice¡¯s and Bob¡¯s agent because all entire relays in the network are trustful to each other.}
\end{figure}

The p-bit commitment scheme combing with BB84 protocol has three phases:

\subsection{Preparation Phase}
each of the two relays generate the secret key by using the bit commitment scheme combining with BB84 protocol. The protocol itself begins when Bob sends to Alice weak coherent pulses random prepared in either horizontal, vertical, diagonal or antidiagonal polarized states. After receiving the signals, Alice uses a public channel to notify Bob which signals she has detected.
\subsection{Commitment Phase}
Alice chooses random basis to measure the signal which she has received. We definite that 4N signals which Alice has measured are one frame. When Alice uses 2N rectilinear polarized basis and 2N diagonal polarized basis to measure the incoming signals in a frame, we take this frame as the bit commitment frame. Otherwise, we use normal BB84 protocol in the frame. We denote the measurement outcome sequence $S_{2N}^x$ to be the sequence we choose to commit with length 2N and the number of such sequence x. The sequence x consists of N bit 0 and N bit 1. Then we encode the outcome sequence again to get the new sequence $S_{2N}^x0$ and $S_{2N}^x1$ to decrease the bit of the measurement outcomes.  We can use probability p to denote the commit probability. To commit value 0 in one frame, Alice chooses the measurement outcome sequence $S_{2N}^x0$ measured in rectilinear polarization basis in a frame. To commit value 1 in one frame, she chooses the measurement outcome sequence $S_{2N}^x1$ measured in diagonal polarization basis in a frame. Then Alice encrypts the measurement sequence with the secret key generated by the new protocol and sends to near the two relays $P_0$ and $P_1$.
\subsection{Unveiling Phase}
	To unveil the commitment, $P_0$ and $P_1$ decrypt the measurement sequence and wait for a moment to verify them. (The time they should wait depends on the average lengths of the two relays and the commitment time required of the whole system. In practice we set each channel a different waiting time to make sure that each one frame ends at the same time in the protocol.) Then Alice discloses the measurement basis the same as in the BB84 protocol. At the same time to verify the commitment, Bob compares the result submitted by $P_0$ and $P_1$. If they are different, Bob rejects the commitment. Otherwise, he estimates a lower bound $n_{rector}$ $n_{diag}$, the number of single photons which sent in the rectilinear or diagonal basis and detected by Alice in the same basis. Let $n_{\epsilon,rect}$ or $n_{\epsilon,diag}$ be the total number of errors in the rectilinear or the diagonal basis. Only when both $n_{rect},n_{diag}\geq N_{tol}$ and $n_{\epsilon,rect}\leq E_{tol}N_{tol}$ or $n_{\epsilon,diag}\leq E_{tol}N_{tol}$ does Bob accept the commitment value as 0 or 1.

The advantage of this p-bit commitment scheme combining with BB84 protocol is that we can use the secret key generated by the system itself to encrypt its measurement outcomes. Depending on the channel error tolerance $Q_{tol}$ , we can get some redundant secret keys to encrypt other information by adjusting an appropriate p which satisfies:
\begin{equation}
p=\frac{C_{4N}^{2N}C_{2N}^N}{2^{4N}2^{2N}\frac{x}{C_{2N}^N}}=\frac{xC_{4N}^{2N}}{2^{4N+2N}}
\end{equation}
where $x\leq C_{2N}^N$

In theoretical, when the channel error tolerance is $Q_{tol}$, the rate of the redundant secret key in a frame satisfies:
\begin{equation}
r'=r-p+\frac{p\log p}{2N}
\end{equation}

\begin{figure}
\includegraphics[width=0.4\textwidth,height=0.3\textheight]{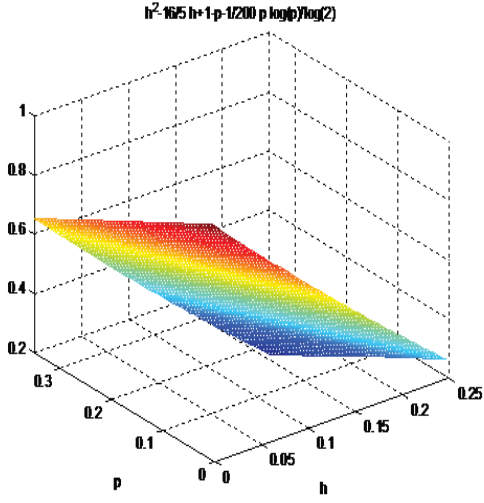}
\caption{The generation rate of the redundant secret key in a frame depends on the channel error tolerance $Q_{tol}$ and the probability p we choose to commitment when we set N=100. }
\end{figure}

By Minkowski causality, $P_0$ represents the probability that Alice unveil data consistent with a 0 commitment at $P_0$. It only depends on the operation that Alice carries out on the line ${AP}_0$. In the same way, we can get $p_1$. It is $\epsilon_b$ binding if Bob has a guarantee that $p_0+p_1\leq 1+\epsilon_b$.

Security against a dishonest Bob: The protocol is easily proved to be secure against Bob, who knows nothing about the information from Alice. Because Alice selects the measurement basis which is independent of the detecting signal.

Security against a dishonest Alice: It states that the bit commitment protocol considered in the paper is $\epsilon_b$ binding, with $\epsilon_b$ approximating zero when $N_{tol}$ increases, given that the tolerated value $E_{tol}$ is not too large. We can get the $\epsilon_b$ binding in this paper:
\begin{gather}
\epsilon_b<p2^{h(p)}*\inf_{\delta\in(E_{tol},\frac{1}{2})}\{[1-\exp(\frac{(\delta N_{tol}-\lfloor E_{tol}N_{tol})^2}{1-N_{tol}})] \notag \\
2^{1-[1-h(\delta)]N_{tol}}+2\exp(\frac{(\delta N_{tol}-\lfloor E_{tol}N_{tol})^2}{1-N_{tol}})\}\notag \\
[1+\sum_{k=1}^{\lfloor E_{tol}N_{tol}\rfloor}(2^k-1)C_{N_tol}^k]
\end{gather}
where $p=\frac{xC_{4N}^{2N}}{2^{4N+2N}}$

\section{Application}
We can use this improved quantum commitment scheme combining with BB84 protocol to the large-scale trustful quantum relay network routing method. Depending on the every channel¡¯s secret key buffer, we can use method of bit commitment to build a better routing method. When we use packet switching, we should choose the method of either datagram or virtual circuit. However, when we choose virtual circuit, we cannot control the waste of information between the source and some other routes which have not been chosen. So we proposed a new quantum relay network routing method based on the improved quantum commitment scheme combining with BB84 protocol.
\subsection{Preparation}
The packet with the length L is needed to be transmitted from relay A to relay B. (There is no other relays in the middle of A and B.) We should consider the buffer $b_{AB}$ in the channel between A and B. If we have n packets to be transmitted from A to B, we can get the probability $P_{AB}$ that one of n packets can be served.
\begin{equation}
P_{AB} = \left\{
             \begin{array}{lcl}
             {\frac{b_{AB}}{nL}   ~~~~~~~~~~   b_{AB}<nL }\\
             {1                   ~~~~~~~~~~~~~~   b_{AB}\geq nL}
             \end{array}
        \right.
\end {equation}

\subsection{Datagram}
When using the method of datagram, we can utilize flooding request to get all possible path that each packet can pass through. Then each relay route gets the information how many packets may pass from it. So the route can provide the serve probability to the next hop depending on their channel secret key buffer. The source compares the probability of every possible path to find the maximum probability way to transmit the packet.

\subsection{Virtual Circuit}
When using the method of virtual circuit, we also utilize flooding request to get all possible path that each packet can pass through. We balance the served probability of each channel and the minimum relays each path has to find a better virtual circuit. After choosing the virtual circuit path, we use the improved quantum commitment combining with BB84 protocol to commit to all relays which path we will use, so that these relays can abort the future mission and provide more secret keys buffer to serve the packet fluently.

The advantage of this method is that when we make sure which path we will choose, we can commit to the entire relay route in this path instead of influence on others. In virtual circuit, if we find the path we will transmit, we should tell the entire routes in the circuit the information transmitted through. And the eavesdropper can be aware of the circuit easily. However, when we use quantum bit commitment, we can use BB84 protocol to accomplish the commitment. It is concealing that the source does not transmit the actual information to the next route and the next route just recognize that the source wants to transmit information. It is binding that after committing to transform, the source cannot change circuit. It can save a lot of buffer that the source does not need to commit with other routes.

\section{Conlusion}
The protocol we proposed in this paper can be easily used in the quantum key distribution systems. So we do not need to build an independently quantum bit commitment system cooperating with other quantum key distribution systems. So it is low in cost. And from the rate of the generation of the redundant secret key we can also get the secret keys to encrypt the information by choosing an appropriate probability p-bit commitment.

\begin{acknowledgments}
This work was supported by the National Natural Science Foundation of China No.61372076 and No.61301171.
\end{acknowledgments}



%
%

%





\begin{thebibliography}{1}
\bibitem{Blum81}
M.~Blum, ``in Advances in Cryptology: A report on CRYPTO'81,'' Santa Barbara, Califomia, USA,  (1981).

\bibitem{Blum82}
M.~Blum, "in Proceedings of the 24th IEEE Computer Conference,"  New York, pp.133-137, (1982)

\bibitem{Brassard88}
G.~Brassard, "Modern Cryptology: A Tutorial, Lecture Notes in Computer Science," Springer-Berlag, New York, pp.325, (1988)

\bibitem{Yao94}
A.~Yao, "in Proceedings of the 26th Annual ACM Symposium on the Theory of Computing," Montreal, p. 67, (1994)

\bibitem{kilian91}
J.~Kilian, "in Proceedings of the 20th Annual ACM Symposium on the Theory of Computing," Montreal, p. 20, (1991)

\bibitem{Mayers97}
D. Mayers "Unconditionally secure quantum bit commitment is impossible," Phys. Rev. Lett. 78, 3414 (1997)

\bibitem{Kwong97}
Hoi-Kwong ~Lo and H. ~F. ~Chau, "Is quantum bit commitment really possible?" Phys. Rev. Lett. 78, 3410 (1997)

\bibitem{Kwong98}
Hoi-Kwong ~Lo and H. ~F. ~Chau, "Why quantum bit commitment and ideal quantum coin tossing are impossible." Physica D120, 177-187 (1998)

\bibitem{Kent99}
A. ~Kent, "Unconditionally secure bit commitment by transmitting measurement outcomes." Phys.Rev.Lett. 83, 1447-1450 (1999)

\bibitem{Croke2012}
S. ~Croke and A. ~kent, "Security details for bit commitment by transmitting measurement outcomes." Phys. Rev. A 86, 052309 (2012)

\bibitem{Kent2012}
A. ~Kent, "Quantum tasks in Minkowski space." Class. Quantum Grav. 29 224013 (2012)

\bibitem{Kent2013}
A. ~Kent, "A no-summoning theorem in relativistic quantum theory." Quantum Information Processing 12 (2) 1023-1032 (2013)

\bibitem{Palmer2013}
M. ~C. ~Palmer, "Relativistic quantum information theory and quantum reference frames." arXiv:1312.4780 (2013)

\bibitem{Liu2014}
Y. ~Liu, Y. ~Cao, M. ~Curty, S. ~K. ~Liao, J. ~Wang, K. ~Cui, Y. ~H. ~Li, Z. ~H. ~Lin, Q. ~C. ~Sun, D. ~D. ~Li H. ~F. ~ Zhang, Y. ~Zhao, C. ~Z. ~Peng, Q. ~Zhang, A. ~Cabello and J. ~W. ~Pan£¬¡° Experimental unconditionally secure bit commitment.¡± Phys. Rev. Lett. 112, 010504 (2014)

\bibitem{Lunghi}
T. ~Lunghi, J. ~Kaniewski, F. ~Bussieres, R. ~Houlmann, M. ~Tomamichel, A. ~kent, N. ~Gisin, S. ~Wehner and H. ~Zbinden,"Experimental bit commitment based on quantum communication and special relativity." Phys. Rev. Lett. 111, 180504 (2013)

\bibitem{Scarani2008}
V. ~Scarani and R. ~Renner, "Quantum cryptography with finite resources: unconditional security bound for discrete-variable protocols with one-way postprocessing," Phys. Rev.Lett. 100,200501 (2008)

\bibitem{Tomamichel}
M. ~Tomamichel, C. ~C. ~W. ~Lim, N. ~Gisin and R.~Renner, "Tight finite-key analysis for quantum cryptography," Nat. Commun. 3, 634 (2012)
\end{thebibliography}
\end{document}